\documentclass[12pt,english]{aastex6}
\usepackage[T1]{fontenc}
\usepackage[latin9]{inputenc}
\usepackage{geometry}
\usepackage{mathtools}
\usepackage{amsbsy}
\usepackage{amssymb}
\usepackage{mathdots}
\usepackage{graphicx}
\usepackage{esint}

\makeatletter

\providecommand{\tabularnewline}{\\}

\usepackage{babel}

\makeatother

\usepackage{babel}
\begin{document}

\title{Generation of large-scale magnetic field in convective full-sphere
cross-helicity dynamo}

\author{V.V. Pipin$^{1}$ and N. Yokoi$^{2}$}

\affil{ $^{1}$Institute of Solar-Terrestrial Physics, Russian Academy of
Sciences, Irkutsk, 664033, Russia,\\
 $^{2}$Institute of Industrial Science University of Tokyo, Bw-505,
4-6-1, Komaba, Meguro-ku Tokyo 153-8505, Japan}
\begin{abstract}
We study effects of the cross-helicity in the full-sphere large-scale
mean-field dynamo models of the $\mathrm{0.3M_{\odot}}$ star rotating
with the period of 10 days. In exploring several dynamo scenarios
which are stemming from the cross-helicity generation effect, we found
that the cross-helicity provide the natural generation mechanisms
for the large-scale scale axisymmetric and non-axisymmetric magnetic
field. Therefore the rotating stars with convective envelope can produce
the large-scale magnetic field generated solely due to the turbulent
cross-helicity effect (we call it $\gamma^{2}$-dynamo). Using mean-field
models we compare properties of the large-scale magnetic field organization
that stems from dynamo mechanisms based on the kinetic (associated
with the $\alpha^{2}$ dynamos) and cross-helicity. For the fully
convective stars, both generation mechanisms can maintain large-scale
dynamos even for the solid body rotation law inside the star. The
non-axisymmetric magnetic configurations become preferable when the
cross-helicity and the $\alpha$-effect operate independently of each
other. This corresponds to situations of the purely $\gamma^{2}$
or $\alpha^{2}$ dynamos. Combination of these scenarios, i.e., the
$\gamma^{2}\alpha^{2}$ dynamo can generate preferably axisymmetric,
a dipole-like magnetic field of strength several kGs. Thus we found
a new dynamo scenario which is able to generate the axisymmetric magnetic
field even in the case of the solid body rotation of the star. We
discuss the possible applications of our findings to stellar observations. 
\end{abstract}

\section{Introduction}

It is widely accepted that magnetic activity of the late-type stars
is due to the large-scale hydromagnetic dynamo which results from
actions of differential rotation and cyclonic turbulent motions in
their convective envelopes \citep{park,KR80}. Solar and stellar observations
show that the surface magnetic activity forms a complicated multi-scale
structure \citep{2009ARAA_donat,2013AARv2166S}. The large-scale organization
of the surface magnetic activity on the Sun and other late-type stars
could be related to starspots \citep{2005LRSP2.8B}. Currently, there
is no consistent theory simultaneously explaining the large-scale
magnetic activity of the Sun and emergence of sunspots at the solar
photosphere. However, each of these two phenomena can be modeled separately.
Moreover, there is no consensus about details of the origin mechanisms
of the large-scale magnetic activity of the Sun and solar-type stars.
The models of the flux-transport dynamos and the concurrent paradigm
of the distributed turbulent dynamos are outlined in reviews of \citet{chrev05},
\citet{brsu05} and \citet{pi13r}. The origin and formations of sun/star
spots are extensively studied as well, (see, e.g., \citealt{2010ApJ720.233C,2010ApJ719.307K,2012ApJ753L13S,2013ApJ777L.37W}).

Results of direct numerical simulations of solar-type stars and M-dwarfs
\citep{2008ApJ676.1262B,brown2011,2015ApJ813L31Y,2016ApJ819.104G,2018AA609A..51W}
show that magnetic field and turbulent convective flows are highly
aligned at the near surface layers. Generally, it is found that in
the regions occupied by the magnetic field the cross-helicity density
$\left\langle \gamma\right\rangle =\left\langle \mathbf{u}\cdot\mathbf{b}\right\rangle $
is not zero. Here, $\mathbf{u}$ and $\mathbf{b}$ are the convective
velocity and fluctuating magnetic field. Alignment of the velocity
and magnetic field is found inside sunspots \citep{2011JPhCS.271a2001B}.
Analysis of the full-disk solar magnetograms show that similar to
the current helicity \citep{zetal10}, the cross-helicity can have
the hemispheric rule \citep{2011SoPh..tmp...39Z}. In other words,
the signs of $\left\langle \gamma\right\rangle $ in the North and
South hemispheres can be opposite. The spottiness (or spot filling
factor) and magnetic filling factors of the fast rotating solar analogs
are estimated to be much larger than for the Sun \citep{2005LRSP2.8B}.
The same is true for the fully convective stars. However physical
properties of starspots may change with a decrease of the stellar
mass (see the above-cited review). Results of the stellar magnetic
cartography (the so-called ZDI methods) showed that the fast rotating
M-dwarfs demonstrate the strong large-scale dipole (or multi-pole)
poloidal magnetic field of strength $>1$kG \citep{2008MNRAS390.567M}.
Using the solar analogy we could imagine that such a field can be
accompanied by the cross-helicity density magnitude which is observed
in sunspots. This leads to a question about how the cross-helicity
can affect the large-scale dynamo on these objects.

After \citet{KR80} there was understood that the alignment of the
turbulent convective velocity and the magnetic field is typical for
saturation stage of the turbulent generation due to the mean electromotive
force (EMF), $\boldsymbol{\mathcal{E}}=\left\langle \mathbf{u}\times\mathbf{b}\right\rangle $.
This consideration does not account effects of cross-helicity that
take place in the strongly stratified subsurface layers of the stellar
convective envelope. The direct numerical simulations show the directional
alignment of the velocity and magnetic field fluctuations in the presence
of gradients of either pressure or kinetic energy \citep{2008PhRvL.100h5003M}.
The dynamo scenarios based on the cross-helicity were suggested earlier
in a number of papers \citep{1993ApJ407.540Y,2000ApJ...537.1039Y,2013GApFD107.114Y}

In the current framework of dynamo studies, the effects of non-uniform
large-scale flows are taken into account only through the differential
rotation or so-called $\Omega$-effect {[}$\nabla\times({\bf {U}}\times{\bf {B}})${]}.
In marked contrast to this differential rotation effect, the non-uniform
flow effect has not been considered in the turbulence effects on the
mean-field induction represented by the turbulent electromotive force
$\boldsymbol{\mathcal{E}}=\left\langle \mathbf{u}\times\mathbf{b}\right\rangle $
. However, if we see the velocity and magnetic-field fluctuation equations
in the presence of the inhomogeneous mean velocity, the turbulent
cross helicity, defined by the correlation between the velocity and
magnetic-field fluctuations ($\left\langle \mathbf{u}\cdot\mathbf{b}\right\rangle $),
should naturally enter the expression of the turbulent EMF as the
coupling coefficient for the mean absolute vorticity (rotation and
the mean relative vorticity) \citep{2016ApJ824.67Y}. This suggests
that, in the presence of a non-uniform large-scale flow, the turbulent
dynamo mechanism arising from the cross helicity should be taken into
account as well as the counterparts of the turbulent magnetic diffusivity
and turbulent helicity or the so-called $\alpha$-effect.

\begin{figure}
\includegraphics[width=0.99\columnwidth]{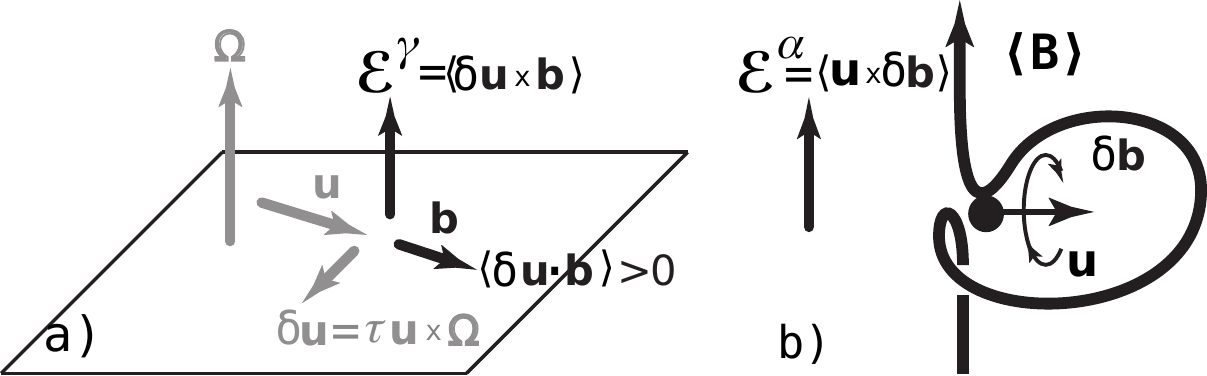}

\caption{a) Generation of the mean electromotive force from cross-helicity
(see explanation in the text); b) the $\alpha$-effect from the axial-aligned
magnetic field.}

\end{figure}

{The cross-helicity may have a particular interest for consideration
of the dynamo mechanisms operating in the fully-convective stars.
For convenience, we briefly remind the physical mechanism behind the
cross-helicity generation effect. Illustration of this mechanism is
shown in Figure 1a. We consider the aligned turbulent velocity and
magnetic field in the plane that is perpendicular to the rotation
axis. The Coriolis force acting on the turbulent motion results in
the mean electromotive force along the rotation axis, $\boldsymbol{\mathcal{E}}^{\gamma}=\left\langle \delta\mathbf{u}\times\mathbf{b}\right\rangle $
(see, Fig1a). This electromotive force can generate the large-scale
toroidal magnetic field due to the $\boldsymbol{\nabla}\times\boldsymbol{\mathcal{E}}^{\gamma}$-term
in the induction equation. The alternative mechanism is given by the
$\alpha$-effect, (see Figure 1b). In this case, the mean electromotive
force results from the large-scale axial magnetic field and the cyclonic
motions in the plane perpendicular to the rotation axis. Note that
for a regime of the fast rotation the energy of the turbulent vortexes
across rotation axis is suppressed. Therefore the axisymmetric $\alpha^{2}$-dynamo
cannot use the axial magnetic field for the dynamo generation \citep{kit-rud:1993b}.
Our consideration gives an idea that the cross-helicity effect can
generate the axisymmetric magnetic field even in the case of the quasi
2D turbulence that is expected for the fast-rotating M-dwarfs. The
differential rotation of the fast rotating M-dwarfs is rather small
\citep{D2-2008MNRAS,2008MNRAS390.567M}. The direct numerical simulations
of \citet{2008ApJ676.1262B} show absence of the differential rotation
in the magnetic case. Therefore we can expect that the axisymmetric
magnetic field is likely generated by means of the turbulent mechanisms
with no regards for the large-scale shear flow. As was mentioned above
the $\alpha$-effect is unlikely to support the axisymmetric dynamo
for the case of the fast-rotating M-dwarfs.}

In this paper, the cross-helicity effects are studied for in the full-sphere
non-axisymmetric dynamo. We identify the different dynamo scenarios
and study the magnetic properties of the fully convective stars using
the nonlinear dynamo models.

\section{Basic equations}

The mean-field convective dynamo is governed by the induction equation
of the large-scale magnetic field $\mathbf{B}$,

\begin{equation}
\frac{\partial\mathbf{B}}{\partial t}=\nabla\times\left(\mathbf{\mathcal{E}}+\overline{\boldsymbol{U}}\times\left\langle \mathbf{B}\right\rangle \right),\label{eq:dyn}
\end{equation}
where $\mathbf{\mathcal{E}}=\left\langle \mathbf{u}\times\mathbf{b}\right\rangle $
is the mean electromotive force of the turbulent convective flows,
$\mathbf{u}$, and, the turbulent magnetic field, $\mathbf{b}$. The
mean-electromotive force includes the generation effects due to the
helical turbulent flows and magnetic field, the generation effect
due to the cross-helicity. {Also, it includes the turbulent
pumping, and the anisotropic (because of rotation) eddy-diffusivity
etc.. For convenience, we divide the expression of the mean-electromotive
force into two parts:
\begin{eqnarray}
\boldsymbol{\boldsymbol{\mathcal{E}}} & = & \boldsymbol{\boldsymbol{\mathcal{E}}}^{\alpha,\eta,V}+\boldsymbol{\boldsymbol{\mathcal{E}}}^{\gamma},\label{eq:emf}
\end{eqnarray}
where $\boldsymbol{\boldsymbol{\mathcal{E}}}^{\gamma}$ results from
the cross-helicity effect (\citealp{2013GApFD107.114Y}, hereafter
Y13): 
\begin{equation}
\boldsymbol{\boldsymbol{\mathcal{E}}}^{\gamma}=C_{\gamma}\frac{\boldsymbol{\Omega}}{\Omega}\left\langle \gamma\right\rangle f_{\gamma}\left(\Omega^{*}\right)\psi_{\gamma}\left(\beta\right),
\end{equation}
where, $\left\langle \gamma\right\rangle =\left\langle \mathbf{u}\cdot\mathbf{b}\right\rangle $,
$\Omega^{*}=4\pi\tau_{c}/P^{*}$ is the Coriolis number, where $\tau_{c}$
is the convective turnover time and the $P^{*}$ is the period of
rotation of a star. Analytical calculations of Y13 do not include
the nonlinear feedback due to the Coriolis force and the large-scale
magnetic field. In our study, these effects will be treated in a simplified
way via quenching functions $f_{\gamma}\left(\Omega^{*}\right)$,
and $\psi_{\gamma}\left(\beta\right)$. The function $f_{\gamma}\left(\Omega^{*}\right)$
takes into account the nonlinear effect of the Coriolis force in the
fast rotation regime. The function $\psi_{\gamma}\left(\beta\right)$,
where $\mathrm{\beta=\left\langle \left|\mathbf{B}\right|\right\rangle /\sqrt{4\pi\overline{\rho}u'^{2}}}$,
$\mathrm{u'}$ is the RMS of the convective velocity, describes the
magnetic quenching of the cross-helicity dynamo effect. Those quenching
functions will be specified later.}

{The part $\boldsymbol{\boldsymbol{\mathcal{E}}}^{\alpha,\eta,V}$
includes the others common contributions of the mean-electromotive
force. It is written as follows:}

\begin{eqnarray}
\boldsymbol{\boldsymbol{\mathcal{E}}}^{\alpha,\eta,V} & = & \hat{\alpha}\circ\left\langle \mathbf{B}\right\rangle +\mathbf{V}^{\left(p\right)}\times\mathbf{\left\langle B\right\rangle }\\
 & - & \left(\eta_{T}+2\eta_{T}^{(\parallel)}\right)\boldsymbol{\nabla}\times\mathbf{\left\langle B\right\rangle }-2\eta_{T}^{(\parallel)}\frac{\boldsymbol{\Omega}}{\Omega^{2}}\mathbf{\boldsymbol{\Omega}\cdot\left(\boldsymbol{\nabla}\times\mathbf{\left\langle B\right\rangle }\right)},\nonumber 
\end{eqnarray}

It is convenient to divide the large-scale magnetic field induction
vector for axisymmetric and non-axisymmetric parts as follows: 
\begin{eqnarray}
\left\langle \mathbf{B}\right\rangle  & = & \overline{\mathbf{B}}+\tilde{\mathbf{B}}\label{eq:b0}\\
\mathbf{\overline{B}} & = & \hat{\boldsymbol{\phi}}\mathrm{B}+\nabla\times\left(\frac{\mathrm{A}\hat{\boldsymbol{\phi}}}{r\sin\theta}\right)\label{eq:b1}\\
\tilde{\mathbf{B}} & = & \mathrm{\boldsymbol{\nabla}\times\left(\hat{\mathbf{r}}T\right)+\boldsymbol{\nabla}\times\boldsymbol{\nabla}\times\left(\hat{\mathbf{r}}S\right),}\label{eq:b2}
\end{eqnarray}
where $\overline{\mathbf{B}}$ is the axisymmetric, and $\tilde{\mathbf{B}}$
is non-axisymmetric part of the large-scale magnetic field; $\mathrm{A}$,
$\mathrm{B}$, $\mathrm{T}$ and $\mathrm{S}$ are scalar functions;
$\hat{\boldsymbol{\phi}}$ is the unit vector in the azimuthal direction
and $\hat{\mathbf{r}}$ is the radius vector; $r$ is the radial distance
and $\theta$ is the polar angle. The cross-helicity pseudo-scalar
is decomposed to the axisymmetric and non-axisymmetric parts, as well,
\begin{equation}
\left\langle \gamma\right\rangle =\overline{\gamma}+\tilde{\gamma}.\label{eq:gam}
\end{equation}

For the non-axisymmetric part of the problem we employ the spherical
harmonics decomposition, i.e., the scalar functions $\mathrm{T}$,
$\mathrm{S}$ and $\tilde{\gamma}$ are represented as follows: 
\begin{eqnarray}
\mathrm{T\left(r,\mu,\phi,t\right)} & \mathrm{=} & \mathrm{\sum\hat{T}_{l,m}\left(r,t\right)\bar{P}_{l}^{\left|m\right|}\exp\left(im\phi\right),}\label{eq:tdec}\\
\mathrm{S\left(r,\mu,\phi,t\right)} & \mathrm{=} & \mathrm{\sum\hat{S}_{l,m}\left(r,t\right)\bar{P}_{l}^{\left|m\right|}\exp\left(im\phi\right),}\label{eq:sdec}\\
\tilde{\gamma}\left(r,\mu,\phi,t\right) & = & \sum\hat{\gamma}_{l,m}\left(r,t\right)\bar{P}_{l}^{\left|m\right|}\exp\left(im\phi\right)
\end{eqnarray}
where $\mathrm{\bar{P}_{l}^{m}}$ is the normalized associated Legendre
function of degree $\mathrm{l\ge1}$ and order $\mathrm{m\ge1}$.
Note that $\mathrm{\hat{S}_{l,-m}=\hat{S}_{l,m}^{*}}$ and the same
for $\hat{T}_{l,m}$ and $\hat{\gamma}_{l,m}$.

The equations governing evolution of the axisymmetric part of the
magnetic field are as follows:

\begin{eqnarray}
\frac{\partial B}{\partial t} & = & \frac{1}{r}\left(\frac{\partial r\boldsymbol{\mathcal{E}}_{\theta}^{\alpha,\eta,V}}{\partial r}+\sin\theta\frac{\partial\boldsymbol{\mathcal{E}}_{r}^{\alpha,\eta,V}}{\partial\mu}\right)-\frac{\sin\theta}{r}\frac{\partial\left(\Omega,A\right)}{\partial\left(r,\mu\right)}\label{eq:B}\\
 & - & \frac{1}{r}\left(\frac{\partial r\overline{U}_{r}B}{\partial r}-\sin\theta\frac{\partial\overline{U}_{\theta}B}{\partial\mu}\right)\nonumber \\
 & - & \frac{C_{\gamma}\sin\theta}{r}\left(\frac{\partial}{\partial r}\left(r\left\langle \gamma\psi_{\gamma}\left(\beta\right)\right\rangle f_{\gamma}\left(\Omega^{*}\right)\right)-\frac{\partial}{\partial\mu}\left(\mu\left\langle \gamma\psi_{\gamma}\left(\beta\right)\right\rangle f_{\gamma}\left(\Omega^{*}\right)\right)\right),\nonumber \\
\frac{\partial A}{\partial t} & = & r\sin\theta\boldsymbol{\mathcal{E}}_{\phi}^{\alpha,\eta,V}-\left(\boldsymbol{\overline{U}}\cdot\boldsymbol{\nabla}\right)A,\label{eq:A}
\end{eqnarray}
where all the parts of the $\boldsymbol{\mathcal{E}}$ except the
cross-helicity effect are written in the symbolic form (see, e.g.,
\citealt{2016ApJ824.67Y}). The Eq(\ref{eq:A}) does not have a contribution
from the cross-helicity. This results in a difference in the cross-helicity
dynamo for the axisymmetric and non-axisymmetric magnetic fields.
Another important observation is that the cross-helicity dynamo can
contribute to generation of the axisymmetric toroidal magnetic field
even when the axisymmetric cross-helicity is zero. It results from
condition $\overline{\tilde{\gamma}\psi_{\gamma}\left(\beta\right)}\ne0$
for the nonlinear case in presence of the axisymmetric and non axisymmetric
magnetic field.

To get the equation for the functions $\mathrm{S}$ and $\mathrm{T}$
we follow the procedure, which is described in detail by \citet{KR80}.
For example, to get the equation for the $\mathrm{S}$ we take the
scalar product of the Eq(\ref{eq:dyn}) with the $\hat{\mathrm{\boldsymbol{r}}}$
and for the equation governing the $\mathrm{T}$ we do the same after
taking the curl of the Eq(\ref{eq:dyn}). Therefore we will have,

\begin{eqnarray}
-\frac{\partial\Delta_{\Omega}S}{\partial t} & = & \boldsymbol{\hat{\mathrm{r}}}\cdot\nabla\times\left(\mathbf{\boldsymbol{\mathcal{E}}}+\overline{\boldsymbol{U}}\times\left\langle \mathbf{B}\right\rangle \right),\label{eq:S}\\
-\frac{\partial\Delta_{\Omega}T}{\partial t} & = & \boldsymbol{\hat{\mathrm{r}}}\cdot\nabla\times\nabla\times\left(\boldsymbol{\mathbf{\mathcal{E}}}+\overline{\boldsymbol{U}}\times\left\langle \mathbf{B}\right\rangle \right),\label{eq:T}
\end{eqnarray}
where $1/r^{2}\Delta_{\Omega}$ is the Laplace operator on the surface
$r=\mathrm{const}$.

With the contributions of the cross-helicity the Eqs(\ref{eq:S},\ref{eq:T})
are re-written as follows 
\begin{eqnarray}
-\frac{\partial\Delta_{\Omega}S}{\partial t} & = & \boldsymbol{\hat{\mathrm{r}}}\cdot\nabla\times\left(\mathbf{\boldsymbol{\boldsymbol{\mathcal{E}}}^{\alpha,\eta,V}}+\overline{\boldsymbol{U}}\times\left\langle \mathbf{B}\right\rangle \right)+C_{\gamma}f_{\gamma}\left(\Omega^{*}\right)\frac{\partial}{\partial\phi}\left(\tilde{\gamma}\psi_{\gamma}\left(\beta\right)\right),\label{eq:S-1}\\
-\frac{\partial\Delta_{\Omega}T}{\partial t} & = & \boldsymbol{\hat{\mathrm{r}}}\cdot\nabla\times\nabla\times\left(\boldsymbol{\boldsymbol{\mathcal{E}}}^{\alpha,\eta,V}+\overline{\boldsymbol{U}}\times\left\langle \mathbf{B}\right\rangle \right)\label{eq:T-1}\\
 & - & \frac{C_{\gamma}f_{\gamma}\left(\Omega^{*}\right)}{r}\Delta_{\Omega}\left(\mu\tilde{\gamma}\psi_{\gamma}\left(\beta\right)\right)+\frac{C_{\gamma}}{r}\frac{\partial}{\partial\mu}\left(\sin^{2}\theta\frac{\partial}{\partial r}r\tilde{\gamma}f_{\gamma}\left(\Omega^{*}\right)\psi_{\gamma}\left(\beta\right)\right).\nonumber 
\end{eqnarray}
We put some details about derivation of Eqs(\ref{eq:S-1}, \ref{eq:T-1})
in Appendix. From these equations, we see that the nonaxisymmetric
part of the cross-helicity is coupled with the evolution of the non
axisymmetric magnetic field. This can provide the dynamo instability
of the large-scale non axisymmetric magnetic field. In particular,
the nonaxisymmetric magnetic field could be generated solely due to
the cross-helicity dynamo effect.

In general case, all coefficients in the Eq(\ref{eq:emf}) depends
on the Coriolis number $\Omega^{*}=4\pi\tau_{c}/P^{*}$, where $\tau_{c}$
is the convective turnover time and the $P^{*}$is the period of rotation
of a star. Also, the magnetic feedback on the generation and transport
effects in the $\boldsymbol{\boldsymbol{\mathcal{E}}}$ should be
taken into account. For the case of $\Omega^{*}\gg1$, the $\alpha$-effect
tensor can be represented as follows \citep{kit-rud:1993b,pi08Gafd}:
\begin{eqnarray}
\mathrm{\alpha_{ij}} & \mathrm{\approx} & \mathrm{c_{\alpha}u'\ell\left|\mathbf{\boldsymbol{\Lambda}}^{(\rho)}\right|\cos\theta\psi_{\alpha}(\beta)f_{5}^{(a)}\left(\Omega^{*}\right)\left\{ \delta_{ij}-\frac{\Omega_{i}\Omega_{j}}{\Omega^{2}}\right\} }\label{alp2d}
\end{eqnarray}
where, $\mathrm{\mathbf{\boldsymbol{\Lambda}}^{(\rho)}=\boldsymbol{\nabla}\log\overline{\rho}}$
is the gradient of the mean density. Although, $f_{5}^{(a)}\left(\Omega^{*}\right)\rightarrow\pi/2$,
when $\Omega^{*}\gg1$, we will keep the dependence on the Coriolis
number for the nonlinear solution. For the case $\Omega^{*}\gg1$,
the magnitude of the kinetic part $\alpha$ effect is given by 
\begin{equation}
\alpha_{0}={\displaystyle \frac{\pi}{2}c_{\alpha}u'\ell\left|\mathbf{\boldsymbol{\Lambda}}^{(\rho)}\right|}.\label{eq:alpha0}
\end{equation}
The magnetic quenching function of the kinetic part of $\alpha$-effect
is defined by 
\begin{equation}
\psi_{\alpha}=\frac{5}{128\beta^{4}}\left(16\beta^{2}-3-3\left(4\beta^{2}-1\right)\frac{\arctan\left(2\beta\right)}{2\beta}\right),
\end{equation}
where $\mathrm{\beta=\left\langle \left|\mathbf{B}\right|\right\rangle /\sqrt{4\pi\overline{\rho}u'^{2}}}$.
For the cross-helicity dynamo effect we assume that $\psi_{\gamma}=\psi_{\alpha}$.
For the sake of simplicity, we skip the magnetic quenching due to
the magnetic helicity conservation, (cf, \citealt{pip13M}).

The turbulent pumping of the mean-field contains the sum of the contributions
due to the mean density gradient (see,\citealp{pi08Gafd}, hereafter
P08) and the mean-filed magnetic buoyancy \citep{kp93},

\begin{eqnarray}
\mathbf{V}^{\left(p\right)}\times\mathbf{\left\langle B\right\rangle } & \mathrm{=} & 3\eta_{0}f_{1}^{(a)}\left(\Omega^{*}\right)\left(\frac{\left(\Omega\cdot\boldsymbol{\Lambda}^{(\rho)}\right)}{\Omega^{2}}\boldsymbol{\Omega}\times\mathbf{\left\langle B\right\rangle }-\left(\boldsymbol{\Omega}\cdot\mathbf{\left\langle B\right\rangle }\right)\left(\boldsymbol{\Omega}\times\boldsymbol{\Lambda}^{(\rho)}\right)\right)\label{eq:pumpr}\\
 & + & \frac{\alpha_{MLT}u'}{\Gamma_{1}}\beta^{2}K\left(\beta\right)\mathrm{\mathbf{g}}\times\mathbf{\left\langle B\right\rangle }\nonumber 
\end{eqnarray}
where the $\mathrm{\alpha_{MLT}}$ is the parameter of the mixing
length theory, $\Gamma_{1}$ is the adiabatic exponent and the function
$\mathrm{K\left(\beta\right)}$ is defined in \citep{kp93} and $\mathbf{g}$
is the unit vector in the radial direction. When $\Omega^{*}\gg1$,
we have $f_{1}^{(a)}\left(\Omega^{*}\right)\rightarrow\pi/(8\Omega^{\ast})$.
The function of the Coriolis number $f_{1}^{(a)}\left(\Omega^{*}\right)$
is given in P08. Dependence of the eddy-diffusivity coefficients on
the Coriolis number is as follows 
\begin{eqnarray*}
\eta_{T} & = & \frac{\eta_{0}}{\Omega^{*2}}\left(1-\frac{\arctan\Omega^{*}}{\Omega^{*}}\right)\\
\eta_{T}^{\|} & = & \frac{3\eta_{0}}{4\Omega^{*2}}\left(\left(\Omega^{*2}+3\right)\frac{\arctan\Omega^{*}}{\Omega^{*}}-3\right),
\end{eqnarray*}
where the eddy-diffusivity coefficient is defined ${\displaystyle \eta_{0}=\mathrm{\nu_{0}/Pm_{T}}}$
where $\nu_{0}=\mathrm{u}'\ell/3$ is the eddy viscosity. 

The quenching of the cross-helicity dynamo for the fast rotating case
was not studied before. We will assume that the cross-helicity effect
is quenched in the same way as the turbulent diffusivity coefficients,
i.e., we put 
\begin{equation}
f_{\gamma}\left(\Omega^{*}\right)=\frac{1}{\pi}\frac{\arctan\Omega^{*}}{1+\Omega^{*}}\label{eq:gamq}
\end{equation}
The Eq(\ref{eq:gamq}) affect the amplitude of the cross-helicity
effect in the large-scale dynamo.

The evolution of the cross-helicity is govern by the conservation
law 
\begin{equation}
\frac{\partial\left\langle \gamma\right\rangle }{\partial t}=\frac{1}{3\overline{\rho}}\left(\left\langle \mathbf{B}\right\rangle \cdot\nabla\right)\overline{\rho}\left\langle \mathbf{u}^{2}\right\rangle -2\boldsymbol{\boldsymbol{\mathcal{E}}}\cdot\boldsymbol{\Omega}+\eta_{0}\Delta\left\langle \gamma\right\rangle \label{eq:crhg}
\end{equation}
In the stellar conditions, the typical spatial scale of the density
stratification is much less than the spatial scale of the mean magnetic
field. Thus, the first term in the Eq(\ref{eq:crhg}) dominates the
second one. Either rotation-induced anisotropy of the $\alpha$-effect,
the eddy diffusivity, and the pumping do not contribute to the cross-helicity
generation. Substituting the general expression of the mean-electromotive
force into Eq(\ref{eq:crhg}) we get,

\begin{equation}
\frac{\partial\left\langle \gamma\right\rangle }{\partial t}=\frac{\eta_{0}}{\tau_{c}}\left(\left\langle \mathbf{B}\right\rangle \cdot\boldsymbol{\Lambda}^{(\rho)}\right)+2\eta_{T}\boldsymbol{\Omega}\cdot\left(\boldsymbol{\nabla}\times\mathbf{\left\langle B\right\rangle }\right)-\frac{\Omega\alpha_{MLT}u'}{\Gamma_{1}}\beta^{2}K\left(\beta\right)\sin\theta\left\langle B_{\phi}\right\rangle +\eta_{0}\Delta\gamma,\label{eq:crh}
\end{equation}
For the numerical solution, we reduce the equations to the dimensionless
form. The radial distance is measured in the units of the solar radius,
as usual for the stellar astrophysics. Thus, we will have the following
set of parameters, the $\Omega_{\star}$ is the rotation rate of the
star, the $\nu_{0}$ is the magnitude of the eddy viscosity, the parameter
$\mathrm{Pm_{T}}$.

{The boundary conditions are as follows. The cross-helicity
and magnetic field are put to zero at the inner boundary which is
close to the center of the star. For the top, we use the vacuum boundary
conditions for the magnetic field. The boundary condition for the
cross-helicity at the top is unknown, we put the radial derivative
to zero at the top.}

\subsection{The possible dynamo scenarios}

{The possible dynamo scenarios depend upon the magnetic field
generation mechanisms, such as the $\alpha$-effect, the so-called
$\Omega$-effect (associating with the differential rotation) and
the cross-helicity dynamo effect (denoted as the $\gamma$-effect).
Following conventions of the dynamo theory \citep{KR80}, we can identify
the following scenarios: $\alpha^{2}$, $\alpha^{2}\Omega$, $\gamma^{2}$,
$\gamma^{2}\Omega$, $\alpha^{2}\gamma^{2}$, and $\alpha^{2}\gamma^{2}\Omega$.
More scenarios can be found in \citet{KR80}. From the point of view
of this study, the scenarios of $\gamma^{2}$ , $\alpha^{2}\gamma^{2}$,
and $\alpha^{2}\gamma^{2}\Omega$ present particular interest. All
of them depend on the cross-helicity generation governed by the Eq(\ref{eq:crh}).}

{The conceivable scenarios of the cross-helicity dynamos depend
on the cross-helicity generation effect.} The simplest scenario realized
when the cross-helicity is generated from the axial current, e.g.,
the term $2\eta_{T}\boldsymbol{\Omega}\cdot\left(\boldsymbol{\nabla}\times\mathbf{\left\langle B\right\rangle }\right)$
in the RHS of the Eq(\ref{eq:crh}). If we consider the axisymmetric
magnetic field, $\gamma^{2}$ dynamo can give generation of the toroidal
magnetic field from the cross-helicity effect. The poloidal field
is decoupled from the system of the dynamo equation and it can have
only a decaying solution. This scenario was discussed previously by
\citet{1993ApJ407.540Y} and \citet{2013GApFD107.114Y} for the dynamo
in accretion disks.

{In stellar convection zone, the cross-helicity generation
due to the density stratification is one of the most important mechanism.
This is supported by the direct numerical simulations of \citet{2008PhRvL.100h5003M}.
This effect is accounted by the first term in the cross-helicity evolution
equation. With regards to the density stratification, all the dynamo
equations are coupled and there is a possibility for $\gamma^{2}$
dynamo.} In this case, only the non-axisymmetric modes can be unstable
in linear solution because the mean electromotive force $\boldsymbol{\boldsymbol{\mathcal{E}}}=\dots+C_{\gamma}\left\langle \gamma\right\rangle \tau_{c}\boldsymbol{\Omega}+\dots$
has no contribution in the equation for the axisymmetric poloidal
magnetic field (associated with the potential A). In what follows
we discuss $\gamma^{2}$ , $\alpha^{2}\gamma^{2}$, and $\alpha^{2}\gamma^{2}\Omega$
scenarios based on the cross-helicity generation effect which comes
from the first term of the Eq(\ref{eq:crh}).

\section{Results}

\subsection{The eigenvalue problem}

For the linear eigenvalue problem we consider the simplified set of
the equations. We assume that the eddy diffusivity, $\eta_{0}=\nu_{0}/\mathrm{Pm_{T}}$
with $\nu_{0}=5\cdot10^{10}\mathrm{cm^{2}}$/s is constant over radial
distance. This corresponds to set of parameters in our model for the
$0.3\mathrm{M}_{\odot}$ star rotating with the period of 10 days.
The density gradient scale has sharp variation in the upper part of
the star and it is nearly constant in depth. It was found that it
is important to keep the spatial variations of the $\mathbf{\boldsymbol{\Lambda}}_{r}^{(\rho)}$
for the for the cross-helicity evolution equation. For the eigenvalue
problem we introduce a new variable, $\xi=R_{\odot}\mathbf{\boldsymbol{\Lambda}}_{r}^{(\rho)}$,
and employ the adiabatic profile of the density variation scale, 
\begin{equation}
\xi\left(r\right)=\frac{1}{2}R_{\star}R_{\odot}/\left(r\left(R_{\star}-r\right)\right).\label{eq:xi}
\end{equation}
Parameter $\xi$ is nearly uniform in the bulk of the star, having
$\xi\approx-10$ and it rapidly falls to $\xi\approx-500$ toward
the surface. For the sake of simplicity, we put the constant $\xi=-50$
in the pumping terms. The amplitude of the $\alpha$-effect will be
determined by parameter $C_{\alpha}={\displaystyle \alpha_{0}R_{\odot}/\nu_{0}}$.
Additional parameter is the ratio $C_{\tau}={\displaystyle R_{\odot}^{2}/\left(\nu_{0}\tau_{c}\right)}$.
It determines the generation of the cross-helicity. Therefore, in
the linear problem, the reduced expression of the mean-electromotive
force is 
\begin{eqnarray}
\hat{\mathcal{E}}_{i} & = & \mathrm{C_{\alpha}\mathrm{Pm_{T}^{-1}}\left\{ \left\langle B_{i}\right\rangle -\frac{\Omega_{i}\left(\boldsymbol{\Omega}\cdot\left\langle \mathbf{B}\right\rangle \right)}{\Omega^{2}}\right\} }+C_{h}\hat{\gamma}\label{eq:emf-l}\\
 &  & +\mathrm{Pm_{T}^{-1}}\left(\frac{\left(\Omega\cdot\boldsymbol{\xi}\right)}{\Omega^{2}}\boldsymbol{\Omega}\times\mathbf{\left\langle B\right\rangle }-\left(\boldsymbol{\Omega}\cdot\mathbf{\left\langle B\right\rangle }\right)\left(\boldsymbol{\Omega}\times\boldsymbol{\xi}\right)\right),\nonumber \\
 & - & \mathrm{Pm_{T}^{-1}}\left(\left(1+2a\right)\boldsymbol{\nabla}\times\mathbf{\left\langle B\right\rangle }+2a\frac{\boldsymbol{\Omega}}{\Omega^{2}}\mathbf{\boldsymbol{\Omega}\cdot\left(\boldsymbol{\nabla}\times\mathbf{\left\langle B\right\rangle }\right)}\right),\nonumber 
\end{eqnarray}
where $a=\eta_{T}^{\|}/\eta_{T}$. In what follows we assume that
$a=1$. Also, in the linear problem the parameter $C_{h}=C_{\gamma}\Omega_{\star}\tau_{c}$
in the Eq(\ref{eq:emf-l}) absorbs the Coriolis number dependence.
The hat sign in Eq(\ref{eq:emf-l}) means that the mean electromotive
force was scaled about $\nu_{0}$. The cross-helicity is governed
by equation

\begin{equation}
\frac{\partial\hat{\gamma}}{\partial t}=C_{\tau}\mathrm{Pm_{T}^{-1}}\left(\left\langle \mathbf{B}\right\rangle \cdot\boldsymbol{\xi}\right)+\mathrm{Pm_{T}^{-1}}\Delta\hat{\gamma},\label{eq:crh-l}
\end{equation}

\begin{figure*}
\includegraphics[width=1\textwidth]{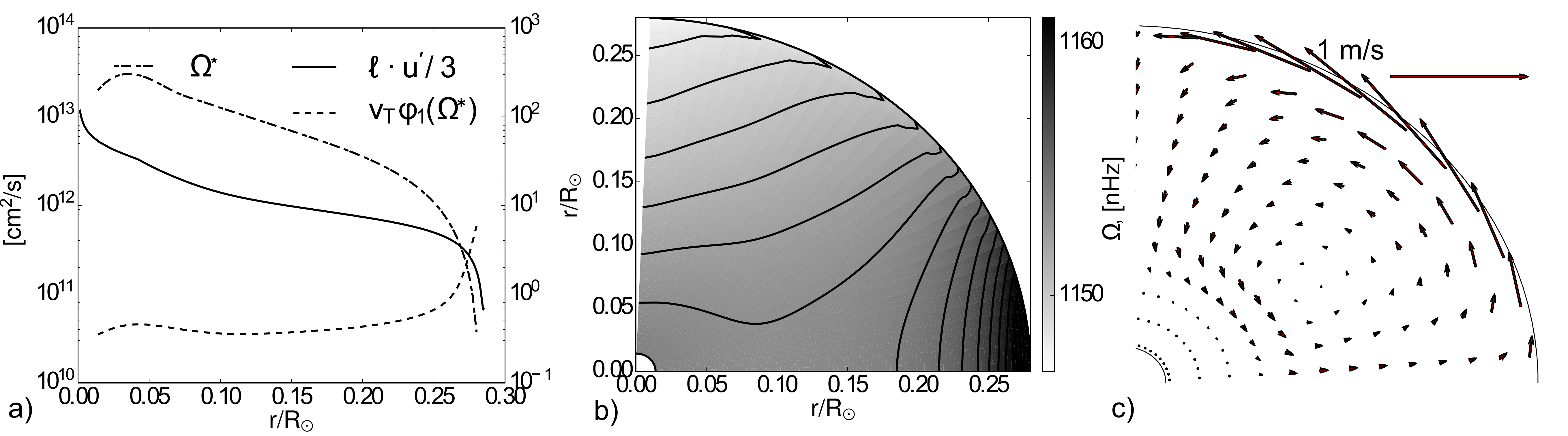} \caption{\label{fig:basic}a) The Coriolis number $\Omega^{*}=2\tau_{c}\Omega_{\star}$
(dash-dot line), where $\tau_{c}$ is the turn-over time of convection,
the turbulent diffusivity parameter, solid line, both the $\tau_{c}$
and the diffusivity are estimated from the MESA code; the dashed line
shows the isotropic eddy viscosity from the heat transport model;
b) angular velocity profiles with contour levels which cover the range
of values which are depicted on the grey scale bar; c) geometry of
the meridional circulation, in the Northern hemisphere.}
\end{figure*}

Our purpose to investigate the eigenvalue solution of the Eqs(\ref{eq:dyn},\ref{eq:emf-l},\ref{eq:crh-l})
for the set of parameters like $\mathrm{Pm_{T}}$ $C_{\alpha}$, $C_{\gamma}$,
and $C_{\tau}$. The effect of the differential rotation can be controlled
by the angular velocity of the star and the distribution of the differential
rotation. We put $C_{\tau}=100$ because the typical diffusive time
scale is order of 100 years and the $\tau_{c}\approx1$ year for this
star. For the external layers of the star, the $C_{\tau}$ is much
larger. We consider the profile of the differential rotation from
our previous paper \citep{2017MNRAS.466.3007P}. It is illustrated
in Figure \ref{fig:basic}. 

{In linear solution, all the partial dynamo modes are decoupled.
We restrict discussion to a few partial modes of the large-scale magnetic
field, including the axisymmetric modes S0 and A0 and the non-axisymmetric
modes S1 and A1. We follow convention suggested by \citet{KR80}:
the letter ``S'' means the mode symmetrical about equator and the
letter ``A'' is for the antisymmetric mode. }

With no regards for the cross-helicity dynamo effect, e.g., in case
of $C_{\tau}=0$ or $C_{h}=0$ , the large-scale dynamo instability
is provided by the $\alpha^{2}$ or $\alpha^{2}\Omega$ scenarios.
For the $\alpha^{2}$ scenario, the critical $C_{\alpha}$ does not
depend on $\mathrm{Pm_{T}}$. Also, in this scenario, the non-axisymmetric
modes are preferable having thresholds at $C_{\alpha}^{(cr)}\approx37$
for the A1 mode and at $C_{\alpha}^{(cr)}\approx42$ for the S1 mode
and for $a=0$. The thresholds are about factor one-half higher for
$a=1$ than for $a=0$.{ This means that the additional diffusive
mixing of the large-scale magnetic field quenches efficiency of the
dynamo mechanisms. In $\alpha^{2}\Omega$ dynamo the instability depends
largely on parameter Pm, because this parameter controls the efficiency
of the large-scale magnetic field stretching by the differential rotation.}
For the $\mathrm{Pm_{T}}$=20, the axisymmetric modes are unstable
first, having thresholds around $C_{\alpha}^{(cr)}\approx10$.

\begin{figure}
\includegraphics[width=1\columnwidth]{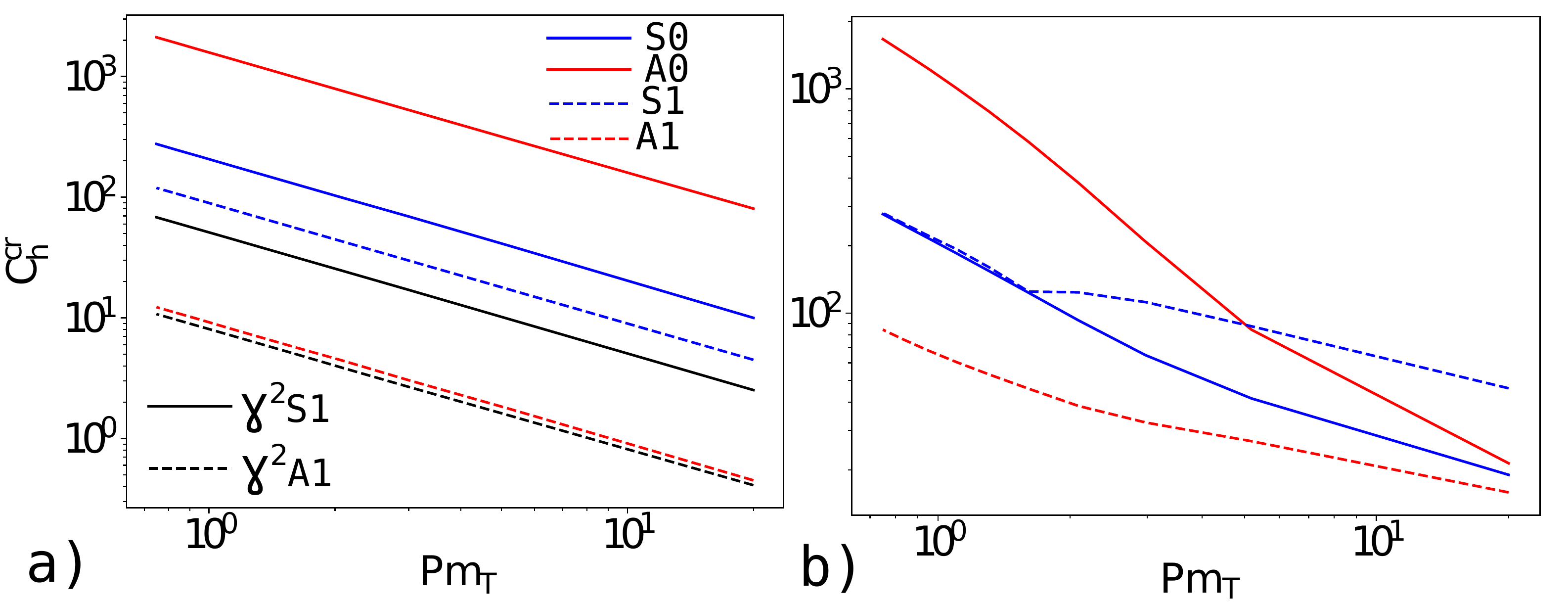}\caption{\label{fig:Ch}a) Excitation thresholds for $\alpha^{2}\gamma^{2}$
scenario (red lines for the A-modes and blue lines for the S-modes)
and $\gamma^{2}$-scenario (black lines, the dashed line is for the
A1 mode and the solid line is for S1-mode) dynamos; b) the same as
(a) for the $\alpha^{2}\Omega\gamma^{2}$ dynamos. The $\alpha$-effect
parameter $C_{\alpha}=10$.}
\end{figure}

The cross-helicity generation effect adds the new parameters for the
study. Results are shown in Figures \ref{fig:Ch}a and b. We restrict
the study by fixing the $C_{\alpha}$ parameter below the dynamo thresholds
of $\alpha^{2}$ and $\alpha^{2}\Omega$ dynamo regimes. {We
put the $C_{\alpha}=10$ and the anisotropy parameter $a=0$, and
study the dynamo instability against the parameter $C_{h}$ for the
variable magnetic Prandtl number $\mathrm{Pm_{T}}$ . Figures \ref{fig:Ch}a
shows results for the pure turbulent dynamo scenarios, i.e., the differential
rotation is disregarded. It is found that the mode A1 keeps the least
dynamo threshold. Also, we found the  $\gamma^{2}$ scenario has the
smaller dynamo instability thresholds for modes A1 and S1 than the
$\alpha^{2}\gamma^{2}$ scenario. Therefore we can conclude that the
magnetic field generation by the concurrent cross-helicity and $\alpha^{2}$
dynamos reduce the efficiency of both dynamo mechanisms. The results
of the linear analysis tells us that without the differential rotation
the nonaxisymmetric dynamo solution is preferable. We can conclude
that during the linear stage of the dynamo process the $\gamma^{2}$
scenario does not give any preference for the axisymmetric magnetic
field generation. Moreover, as it was anticipated from the dynamo
equations the $\gamma^{2}$ scenario provide the additional mechanisms
for the non-axisymmetric magnetic field generation.}

{Figure 2b shows that with an account of the differential rotation
effect, i.e., in considering the $\alpha^{2}\Omega\gamma^{2}$ dynamo
scenario, we get the instability thresholds for all the modes close
to each other with an increase of $\mathrm{Pm_{T}}$. Also, the efficiency
of the axisymmetric dynamo instability increases with the increase
of $\mathrm{Pm_{T}}$. Within the studied parameter range of $\mathrm{Pm_{T}}$
the nonaxisymmetric mode A1 keeps to be the most unstable. We also
studied instability for the spatially nonuniform density stratification
parameter $\xi$. In this case the dynamo instability thresholds are
about a factor of magnitude larger than in the case of the constant
$\xi$. However, the order of the instability thresholds among the
different partial dynamo modes remains the same as it is shown in
Figures \ref{fig:Ch}a and b.}

\subsection{The nonlinear solution}

For the nonlinear solution, we employ the model which keeps the spatial
dependence of the turbulent parameters provided by the {\footnotesize{}{}MESA}
code and solution of the heat transport problem (see Pipin 2017).
Using the results of the eigenvalue problem we bear in mind that the
parameters $C_{h}$ and $C_{\tau}$ in the Eqs(\ref{eq:emf-l}) and
(\ref{eq:crh-l}) absorb the dependence upon the parameter $\tau_{c}$.
{It is found that the Coriolis number parameter $\Omega^{*}=4\pi\tau_{c}/P^{*}$,
where $P^{*}=10$ days, varies from about 1 near the surface to 200
in the depth of the star.} This means that the critical threshold
of the $C_{\gamma}=2C_{h}/\Omega^{*}\ll1\sim0.01$ for $\alpha^{2}\Omega\gamma^{2}$
dynamo if $\mathrm{Pm_{T}}=3$. We use this $\mathrm{Pm_{T}}$ in
all models below. Three different models will be considered. Parameters
of the models are listed in Table 1. The model M1 represent the $\gamma^{2}$
scenario, the model M2 represents the $\alpha^{2}\gamma^{2}$ scenario
and M3 stands for $\alpha^{2}\tilde{\gamma}^{2}$. In the latter case,
we disregard the contribution of the axisymmetric cross-helicity in
the mean-electromotive force. This imitates the situation when the
mean cross-helicity has no hemispheric sign rule. The nonlinear combination
of the nonaxisymmetric magnetic field and cross helicity can produce
the axisymmetric magnetic field. The model M3 was introduced to study
this situation. In this paper, we consider models with the solid body
rotation.

\begin{table}
\caption{}
\centering{}%
\begin{tabular}{|cccc|}
\hline 
\multicolumn{1}{|c|}{Model} & \multicolumn{1}{c|}{Scenario} & \multicolumn{1}{c}{$C_{\gamma}$} & $C_{\alpha}$\tabularnewline
\hline 
M1  & $\gamma^{2}$  & 0.01  & \tabularnewline
\hline 
M2  & $\alpha^{2}\gamma^{2}$  & 0.01  & 0.03\tabularnewline
\hline 
M3  & $\alpha^{2}\tilde{\gamma}^{2}$  & 0.01  & 0.03\tabularnewline
\hline 
\end{tabular}
\end{table}

\begin{figure}
\includegraphics[width=1\columnwidth]{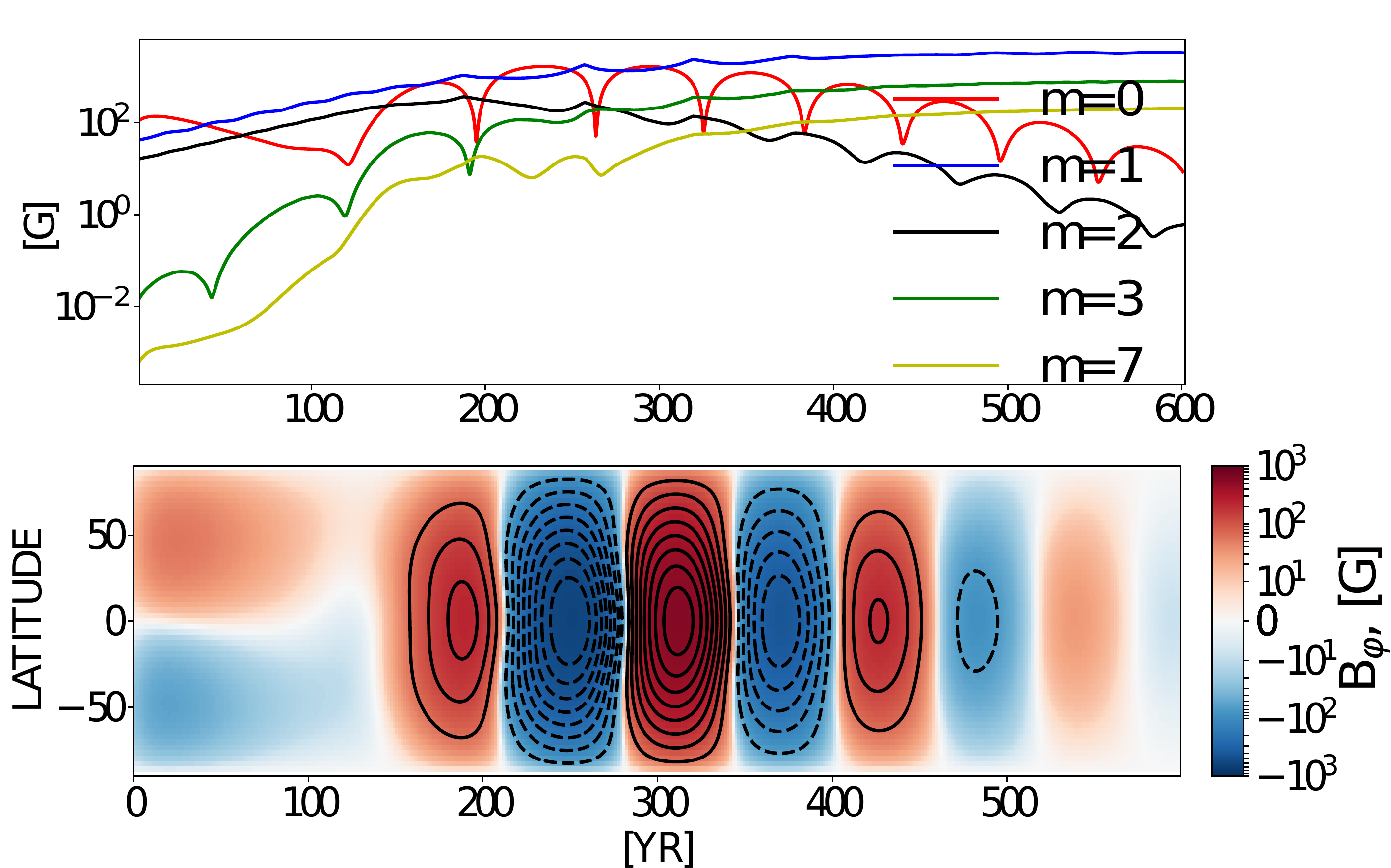}

\caption{\label{M1a}The model M1 ( $\gamma^{2}$ dynamo), a) shows evolution
of the partial modes of the toroidal magnetic field at at the $\mathrm{r=\frac{3}{4}R_{\star}}$;
b) the time-latitude diagram of the toroidal magnetic field shown
by the color image and contours (range of $\pm$1kG). }
\end{figure}

\begin{figure}
\includegraphics[width=1\columnwidth]{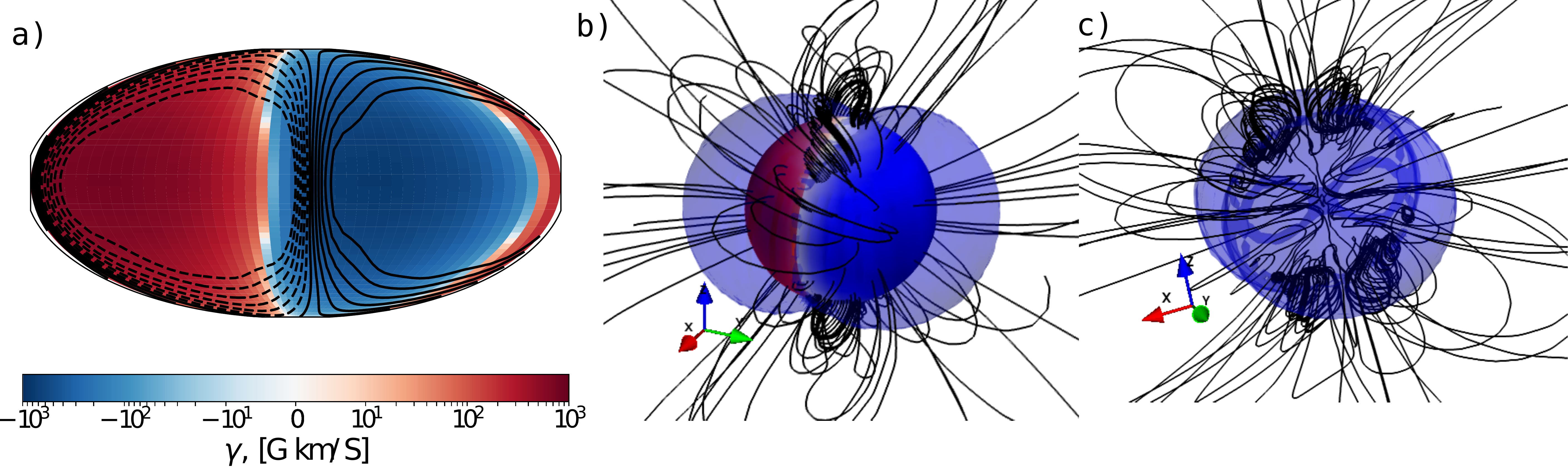}\caption{\label{M1b}The model M1, snapshots of the cross-helicity and magnetic
field distributions in the stationary phase of dynamo evolution, a)
the cross-helicity distributions is shown by color and the radial
magnetic field (iso-lines $\pm1$kG); b) strength of the toroidal
magnetic field at the $\mathrm{r=\frac{3}{4}R_{\star}}$ for axisymmetric
and non-axisymmetric magnetic field and the ratio between energy of
the axisymmetric magnetic field and total magnetic energy; b) the
time-latitude diagram for axisymmetric toroidal magnetic field at
the $\mathrm{r=\frac{3}{4}R_{\star}}$and the color image shows shows
the axisymmetric radial magnetic field at the surface. }
\end{figure}

Regime $\gamma^{2}$ provides the simplest scenario of the cross-helicity
dynamo. Contrary to the $\alpha^{2}$ dynamo, the $\gamma^{2}$ works
only in the nonaxisymmetric regime. In this scenario evolution of
the axisymmetric components of the toroidal and poloidal magnetic
fields are decoupled. Figures \ref{M1a} and \ref{M1b} show evolution
of the partial modes in the model M1 as well as snapshots of the magnetic
field and the cross-helicity distributions at the stationary stage
of evolution. It is seen that the axisymmetric mode of the toroidal
magnetic field evolves non-monotonically showing growth at the beginning
and it decays afterwhile. In nonlinear case, the cross-helicity that
is produced by the non-axisymmetric magnetic field may contribute
to generation of the axisymmetric toroidal magnetic field, because
in general $\left\langle \gamma\psi_{\gamma}\left(\beta\right)\right\rangle \ne0$
(see, eq.\ref{eq:B}). Therefore, the nonaxisymmetric cross-helicity
affects generation of the axisymmetric toroidal magnetic field. However,
evolution of the axisymmetric poloidal magnetic field is decoupled
of the toroidal magnetic field and the axisymmetric cross-helicity.
Therefore, there is no true axisymmetric dynamo in this case. The
axisymmetric field starts to decay when parts of the product $\left\langle \gamma\psi_{\gamma}\left(\beta\right)\right\rangle $
get synchronized. Both the non-axisymmetric cross-helicity density
$\gamma$ and magnetic field can be represented by the equatorial
dipole which changes orientation rotating around the axis of stellar
rotation. This phenomenon is known as the nonaxisymmetric dynamo waves.

\begin{figure}
\includegraphics[width=1\columnwidth]{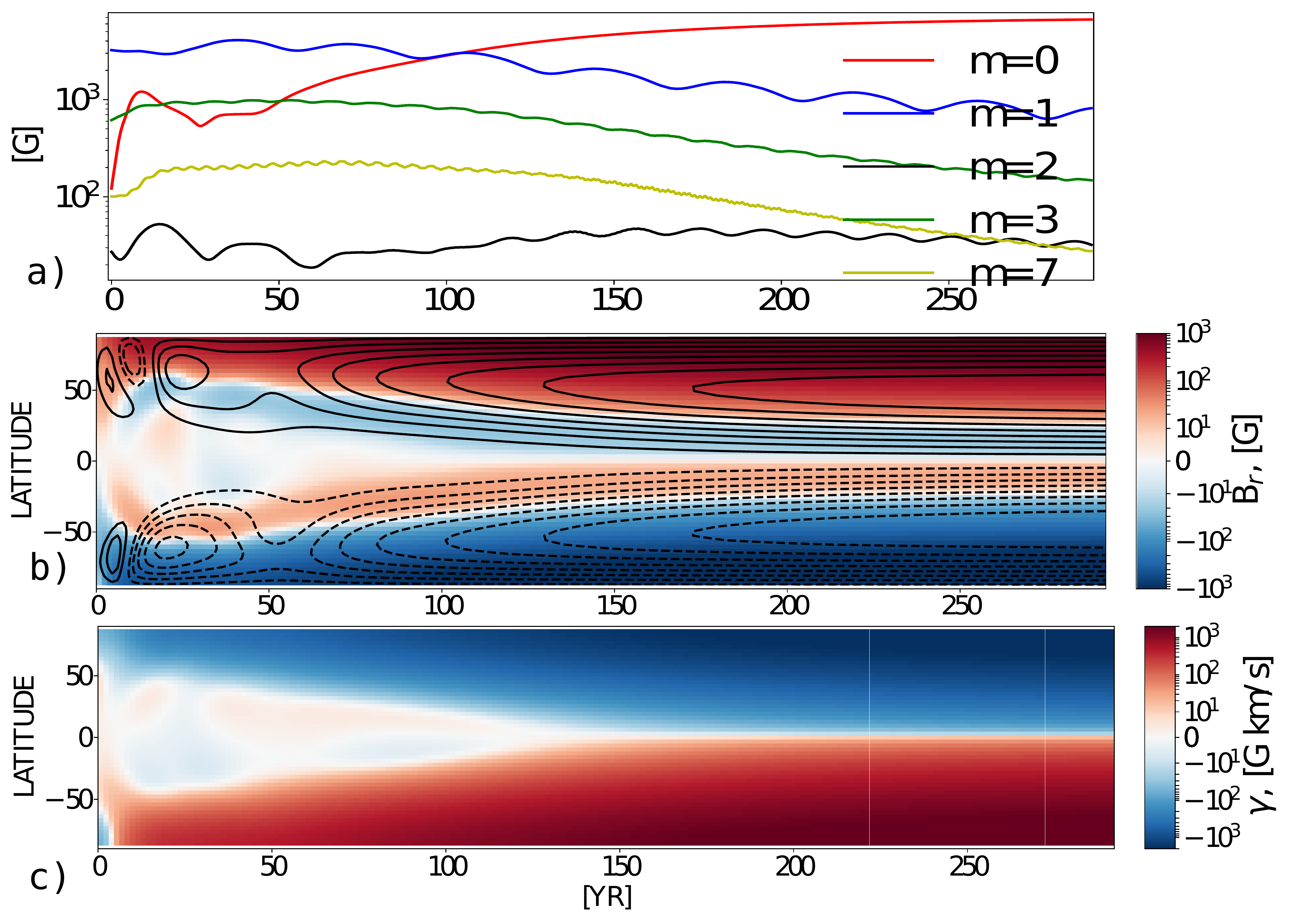}

\caption{\label{M2}The model M2 ($\alpha^{2}\gamma^{2}$ dynamo) a) shows
evolution of the partial modes of the toroidal magnetic field at at
the $\mathrm{r=\frac{3}{4}R_{\star}}$; b) time-latitude diagram of
the radial magnetic field at the surface shown by the color image
and contours show the toroidal magnetic field at at the $\mathrm{r=\frac{3}{4}R_{\star}}$
(range of $\pm$3kG); c) shows the time-latitude evolution for the
cross-helicity}
\end{figure}

\begin{figure}
\includegraphics[width=1\columnwidth]{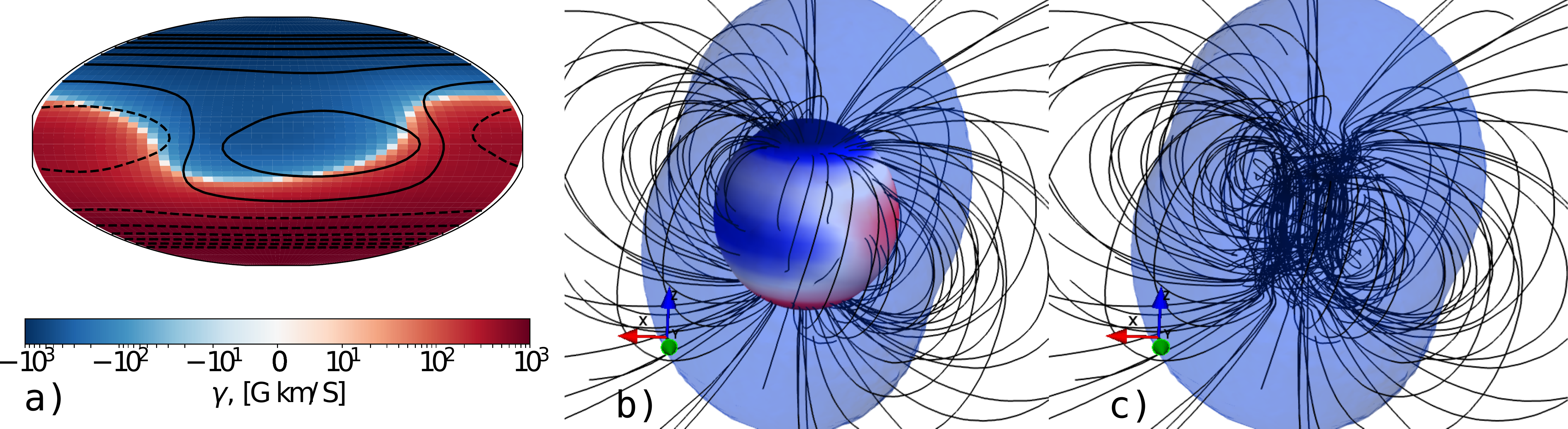}\caption{\label{M2b}The same as Fig.\ref{M1b} for the model M2.}
\end{figure}

The scenario of the $\alpha^{2}\gamma^{2}$ dynamo has a possibility
for the axisymmetric magnetic field generation. Figures \ref{M2}
and \ref{M2b} show evolution of the partial modes in the model M2
as well as snapshots of the magnetic field and the cross-helicity
distributions at the stationary stage of evolution. We use the output
of the model M1 as an initial condition for the model M2. The Figs.\ref{M2}(a)
and (b) show that the axisymmetric toroidal magnetic field started
to grow at the beginning phase showing some oscillations. The dynamo
solution reaches the stage with the constant dipole-like distribution
at the end of simulation. The similar behavior is demonstrated by
the cross-helicity evolution. The cross-helicity has the opposite
signs in the northern and southern hemispheres. The polar magnetic
field in model M2 reaches a magnitude of the 2kG. At the end of simulation,
the model keeps a substantial nonaxisymmetric magnetic field. It has
more than one order of magnitude less strength than the axisymmetric
magnetic field. Snapshots of the magnetic field and cross-helicity
distributions show that these nonaxisymmetric components concentrate
in the near equatorial regions. The field lines of the magnetic field
distribution show that the overall configuration of the magnetic field
is dipole-like both inside and outside the star.

Using the output of the model M2 we made additional run neglecting
the cross-helicity generation effects. This return the dynamo model
to the $\alpha^{2}$ scenario. Similar to \citet{2017MNRAS.466.3007P}
we get the non-axisymmetric magnetic field at the end of the run.
Also, we made additional runs with the decreased $C_{\gamma}$. For
the given parameter $C_{\alpha}$ it was found that the model keeps
axisymmetric magnetic even in the case when $C_{\gamma}$ is by a
factor 2 less than in the model M2.

\begin{figure}
\includegraphics[width=1\columnwidth]{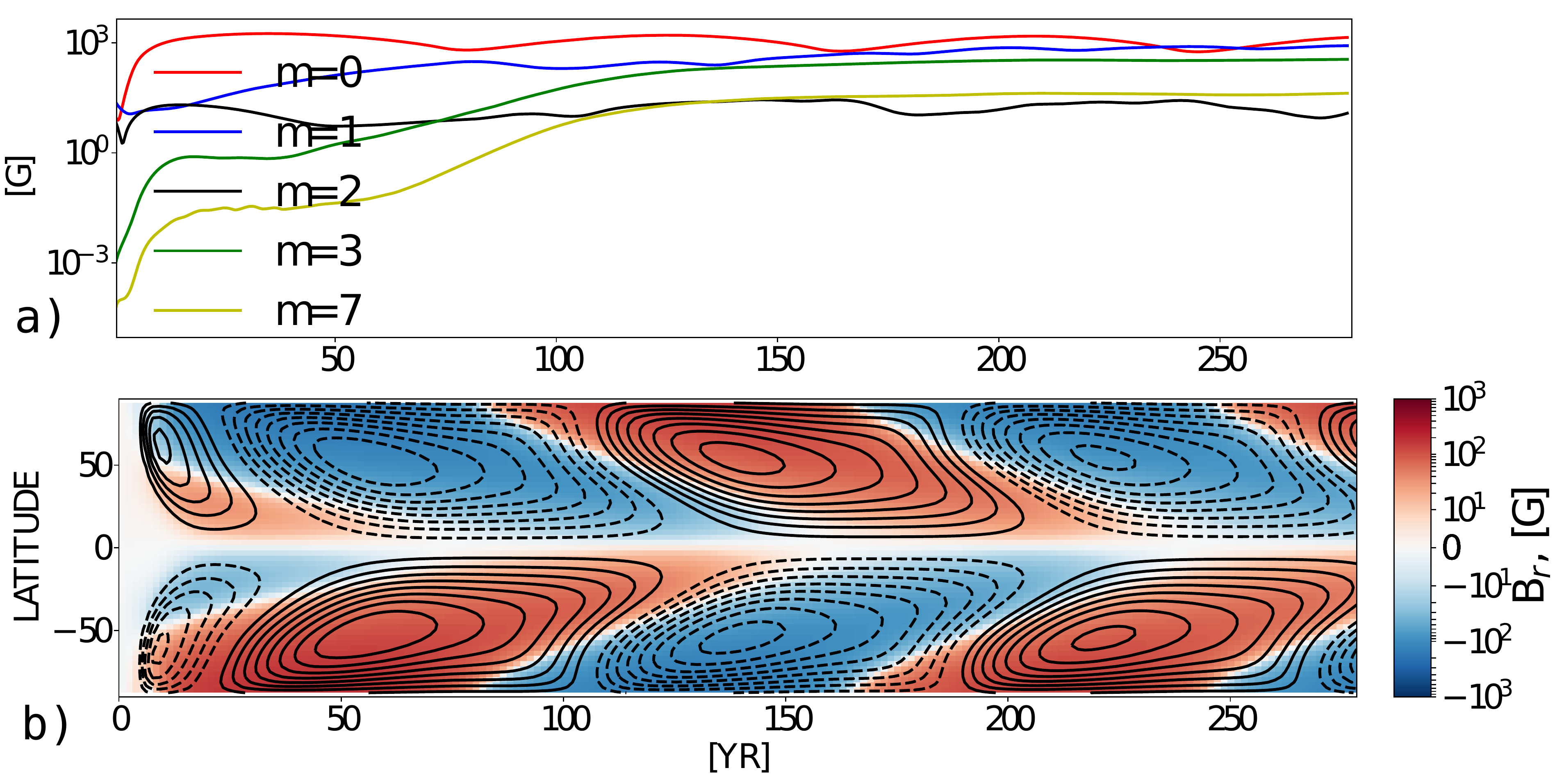}

\caption{\label{M3}The $\alpha^{2}\tilde{\gamma}^{2}$ dynamo, a) shows evolution
of the partial modes of the toroidal magnetic field at at the $\mathrm{r=\frac{3}{4}R_{\star}}$;
b) time-latitude diagram of the radial magnetic field at the surface
shown by the color image and contours show the toroidal magnetic field
at at the $\mathrm{r=\frac{3}{4}R_{\star}}$ (range of $\pm$3kG).}
\end{figure}

\begin{figure}
\includegraphics[width=1\columnwidth]{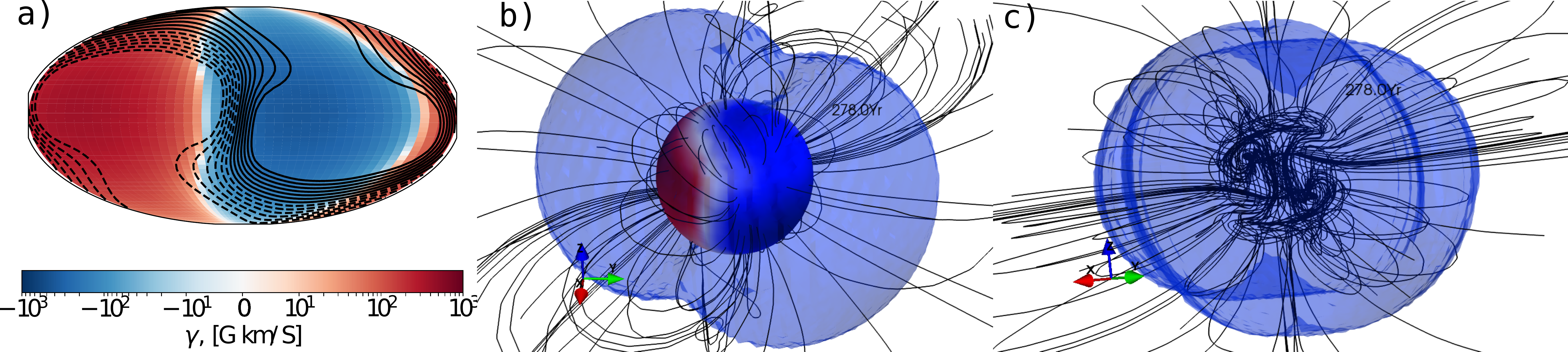}\caption{\label{M3b}The same as Fig.\ref{M1b} for the model M3.}
\end{figure}

The interesting question is if the axisymmetric dynamo can be sustained
by the cross-helicity generation effect when the spatially averaged
cross-helicity is zero. {The model M3 illustrates this scenario.
In this model, we disregard the contribution of the axisymmetric cross-helicity
in the mean-electromotive force by neglecting contributions of the
axisymmetric magnetic field in the equation of the cross-helicity
evolution.} Results are shown in Figures \ref{M3} and \ref{M3b}.
Results show that contrary to the pure $\gamma^{2}$ scenario the
axisymmetric magnetic field is generated. This means that the azimuthally
averaged cross-helicity dynamo effect is not zero, $\left\langle \gamma\psi_{\gamma}\left(\beta\right)\right\rangle \ne0$,
because the terms of the product $\left\langle \gamma\psi_{\gamma}\left(\beta\right)\right\rangle $
are not synchronized in the azimuthal direction. It is caused by the
nonlinear generation of the non axisymmetric magnetic field both by
the $\gamma^{2}$ and the $\alpha^{2}$ mechanisms. We see that the
strength of the axisymmetric and nonaxisymmetric toroidal magnetic
field is same by the order of magnitude. The axisymmetric magnetic
field shows the solar-like time-latitude evolution of the toroidal
magnetic field inside the star. The radial magnetic field at the surface
shows the dominant nonaxisymmetric magnetic field and the nonaxisymmetric
distribution of the cross-helicity. During the nonlinear evolution
the pattern of the magnetic field distribution shown in Fig. \ref{M3b}b
moves about the axis of rotation, representing the azimuthal dynamo
wave. Also, it weakly oscillates around the perpendicular axis which
corresponds to the axis of the equatorial dipole. By this reason,
the model can show the nearly axisymmetric configuration of the polar
magnetic field during the minims and maxims of the axisymmetric magnetic
field cycle. The frequency of the non-axisymmetric m=1 mode is twice
of the axisymmetric one.

\section{Discussion and conclusions}

The physical origin of this cross-helicity effect lies in the combination
of the local angular-momentum conservation in a rotational motion
and the presence of the velocity-magnetic-field correlation \citep{2013GApFD107.114Y}.
Unlike the $\Omega$ effect, the cross-helicity effect does not depend
on the particular configuration of the differential rotation. Provided
that a finite turbulent cross helicity exists, the cross-helicity
effect should work in the presence of the absolute vorticity (rotation
and relative vorticity). This means that we can expect the cross-helicity
dynamo mechanism to work even in the case that the differential rotation
is negligibly small. How and how much cross helicity exists in turbulence
is another problem. In our models, the turbulent cross helicity is
generated by means of the large-scale magnetic field and density stratification.
This generation mechanism was analytically found in a number of papers
\citep{1999PhFl...11.2307Y,2011ApJ...743..160P,2011SoPh..269....3R,2013GApFD107.114Y}
using the mean-field magnetohydrodynamics framework. Our results show
that this turbulent cross-helicity generation effect results in a
number of the new dynamo scenarios.

{It was shown that for the solid body rotation regime there
are three possible dynamo scenarios: the $\gamma^{2}$-dynamo (pure
cross-helicity dynamo), the $\alpha^{2}\gamma^{2}$-dynamo and its
modification - the $\alpha^{2}\tilde{\gamma}^{2}$ dynamo. The latter
is operating from the purely nonaxisymmetric cross-helicity distribution.
In nonlinear case, both the $\gamma^{2}$ and the $\alpha^{2}$ dynamo
scenarios sustain only the nonaxisymmetric magnetic field.} For the
$\gamma^{2}$ scenario the evolution axisymmetric components of the
magnetic field are decoupled. Therefore this regime cannot sustain
the axisymmetric magnetic field against decay. 

{An interesting new effect was considered. It is found that,
in general, it is possible to generate some axisymmetric magnetic
field in nonlinear regime if the spatial variations of the non-axisymmetric
distributions of the cross-helicity and magnetic energy are not synchronized
in the azimuthal direction.} This gives generation of the axisymmetric
toroidal magnetic field due to the cross-helicity dynamo effect because
for the axisymmetric part of $\left\langle \gamma\psi_{\gamma}\left(\beta\right)\right\rangle $
we have 
\begin{equation}
\left\langle \gamma\psi_{\gamma}\left(\beta\right)\right\rangle =\overline{\gamma}\overline{\psi_{\gamma}\left(\beta\right)}+\overline{\tilde{\gamma}\tilde{\psi_{\gamma}\left(\beta\right)}}\ne0,\label{eq:gg}
\end{equation}
where the first term of RHS is zero in $\gamma^{2}$ regime. The second
term of the RHS of this equation is not necessarily zero. Results
of the model M1 show that in the $\gamma^{2}$ scenario the $\tilde{\gamma}$
and $\tilde{\psi_{\gamma}\left(\beta\right)}$ get synchronized. This
prevents the axisymmetric toroidal magnetic field generation. On the
other hand, if there is an additional mechanism of the nonaxisymmetric
magnetic field generation, e.g. by the $\alpha^{2}$ dynamo, then
the axisymmetric dynamo can be excited even if the axisymmetric part
of the cross-helicity is zero. This was demonstrated by model M3 using
the $\alpha^{2}\tilde{\gamma}^{2}$ scenario. There, the effect of
the axisymmetric cross-helicity generation was disregarded. The given
mechanism shows a mixture of the non-axisymmetric and axisymmetric
modes with a dominance of the equatorial dipole-like mode.

One of the most important findings of our work is that the axisymmetric
magnetic field can be generated by means of the cross-helicity generation
and $\alpha^{2}$ in this case we employ the standard formulation
of the mean-electromotive force suggested by \citet{2000ApJ...537.1039Y}.
The dynamo mechanism operates with regards to the axisymmetric and
non-axisymmetric cross-helicity generation. In this case, the strong
axisymmetric dipole-like magnetic field is generated. Unlike the nonlinear
$\alpha^{2}\Omega$ regimes (cf. \citealp{2017MNRAS.466.3007P}),
this scenario produces the constant in time magnetic field configuration
with antisymmetric about equator cross-helicity, toroidal and radial
magnetic field distributions. Our scenario was demonstrated for the
solid body regime. This means that it can be realized on the fast
rotating M-dwarfs with a period of rotation about 1day, which often
show only a small amount of the differential rotation \citep{D2-2008MNRAS}.
Moreover the direct numerical simulation e.g., \citet{2008ApJ676.1262B},
and mean-field models, e.g., \citet{2017MNRAS.466.3007P} show suppression
of the differential rotation in nonlinear regimes. For the solid body
rotation, $\alpha^{2}$ dynamo produce the nonaxisymmetric magnetic
field \citep{2006AA446.1027C,elst07}. This because the $\alpha$-effect
cannot use the component of the large-scale magnetic field along rotation
for generation the axial electromotive force and this results from
the anisotropic $\alpha$-effect in the case of the high Coriolis
number (see, Eq\ref{alp2d} and \citealp{kit-rud:1993b}). The cross-helicity
can generate the poloidal electromotive force in this case \citep{2013GApFD107.114Y}.
This provides generation of the axisymmetric magnetic field by the
$\alpha^{2}\gamma^{2}$-dynamo. The magnetic field configuration produced
in our model of $\alpha^{2}\gamma^{2}$-dynamo is very similar to
those which was found on the fast rotating M-dwarfs, e.g., V374 Peg
and YZ CMi \citep{2008ASPC..384..156D,D2-2008MNRAS,2009ARAA_donat}.

In our paper we employ rather simplified approach to model the cross-helicity
generation effects for the fast rotating regimes. For the further
application the analytical results for the cross-helicity generation
effect in case of the fast rotation and strong magnetic field have
to be developed. We hope that future work could shed more light about
usability of the cross-helicity generation effects in stellar dynamos.

\textbf{Acknowledgments} We thank support RFBR under grant 16-52-50077.
Valery Pipin thank the grant of Visiting Scholar Program supported
by the Research Coordination Committee, National Astronomical Observatory
of Japan (NAOJ) and the project II.16.3.1 of ISTP SB RAS.

%\bibliographystyle{/home/va/work/pap/aastex/3/apj}
%\bibliography{/home/va/work/pap/dyn}

\section{Appendix}

To derive evolution equations for the non-axisymmetric parts of the
magnetic field we use approach suggested by \citet{KR80} and some
useful identities (more of them can be found in their book). For any
scalar functions T and S and radius vector $\hat{\mathbf{r}}$ we
have: 
\begin{eqnarray}
\boldsymbol{\nabla}\times\left(\hat{\mathbf{r}}T\right) & = & -\hat{\mathbf{r}}\times\boldsymbol{\nabla}T\nonumber \\
\boldsymbol{\nabla}\times\boldsymbol{\nabla}\times\left(\hat{\mathbf{r}}S\right) & = & \boldsymbol{\nabla}\frac{\partial rS}{\partial r}-r\Delta S,\nonumber \\
\boldsymbol{\hat{\mathbf{r}}\cdot\nabla}\times\boldsymbol{\nabla}\times\left(\hat{\mathbf{r}}S\right) & = & -\Delta_{\Omega}S\label{eq:delta}\\
 & = & -\frac{1}{\sin\theta}\frac{\partial}{\partial\theta}\left(\sin\theta\frac{\partial S}{\partial\theta}\right)-\frac{1}{\sin^{2}\theta}\frac{\partial^{2}S}{\partial\phi^{2}}\nonumber 
\end{eqnarray}
To derive Eq(\ref{eq:S}), we substitute $\boldsymbol{\mathbf{B}}=\hat{\boldsymbol{\phi}}\mathrm{B}+\nabla\times\left(\frac{\mathrm{A}\hat{\boldsymbol{\phi}}}{r\sin\theta}\right)+\boldsymbol{\nabla}\times\left(\hat{\mathbf{r}}T\right)+\boldsymbol{\nabla}\times\boldsymbol{\nabla}\times\left(\hat{\mathbf{r}}S\right)$
to the LHS of the Eq(\ref{eq:dyn}) and taking into account nonaxisymmetry
and the Eq(\ref{eq:delta}) we get ${\displaystyle \frac{\partial\left(\hat{\mathbf{r}}\cdot\boldsymbol{\mathbf{B}}\right)}{\partial t}=-\frac{\partial\Delta_{\Omega}S}{\partial t}}$.
For the cross-helicity contribution to the RHS of that equation we
have: 
\begin{eqnarray*}
\boldsymbol{\hat{\mathrm{r}}}\cdot\nabla\times\left(\mathbf{\boldsymbol{\mathcal{E}}}^{\gamma}\right) & = & C_{\gamma}\boldsymbol{\hat{\mathrm{r}}}\cdot\nabla\times\left(\frac{\boldsymbol{\Omega}}{\Omega}\left\langle \gamma\right\rangle f_{\gamma}\left(\Omega^{*}\right)\psi_{\gamma}\left(\beta\right)\right)\\
 & = & C_{\gamma}\boldsymbol{\hat{\mathrm{r}}}\cdot\nabla\times\left(\left(\frac{\boldsymbol{\hat{\mathrm{r}}}}{r}\mu-\hat{\theta}\sin\theta\right)\left\langle \gamma\right\rangle f_{\gamma}\left(\Omega^{*}\right)\psi_{\gamma}\left(\beta\right)\right)\\
 & = & C_{\gamma}\nabla\cdot\left(\boldsymbol{\hat{\mathrm{r}}}\times\hat{\theta}\right)\sin\theta\left\langle \gamma\right\rangle f_{\gamma}\left(\Omega^{*}\right)\psi_{\gamma}\left(\beta\right)\\
 & = & C_{\gamma}\frac{\partial}{\partial\phi}\left\langle \gamma\right\rangle f_{\gamma}\left(\Omega^{*}\right)\psi_{\gamma}\left(\beta\right),
\end{eqnarray*}
where $\hat{\theta}$ is the unit vector along polar angle coordinate. 
\end{document}